\documentclass{PoS}
\usepackage{slashed}

\title{Generalized Wandzura Wilczek Relations and Orbital Angular Momentum}

\ShortTitle{Generalized Wandzura Wilczek Relations}

\author{\speaker{Simonetta Liuti}\thanks{Work funded by US DOE Grant DE-SC0016286 and TMD Topical Collaboration}\\
        University of Virginia\\
        E-mail: \email{sl4y@virginia.edu}}

\author{Michael Engelhardt\thanks{Work funded by US DOE Grant DE-FG02-96ER40965 and TMD Topical Collaboration}\\
        New Mexico State University \\
        E-mail: \email{engel@nmsu.edu}}

\author{Abha Rajan\thanks{Work funded by US DOE Grant DE-SC0016286}\\
         University of Virginia\\
        E-mail: \email{arx5c@virginia.edu}}

\abstract{New Lorentz Invariance Relations (LIRs) were presented between twist-three Generalized Parton Distributions (GPDs) and transverse momentum, $k_T$, moments of twist-two Generalized Transverse Momentum-Dependent Distributions (GTMDs). By implementing both these LIRs and the QCD Equations of Motion in the quark quark correlation function, we generated a new series of Wandzura Wilczek (WW) relations in the off-forward sector.  Two of these WW relations take on a particularly interesting physical meaning in that they provide a clear interpretation of the QCD structure of Orbital Angular Momentum (OAM) in the nucleon. In particular, they provide a solution to the outstanding puzzle of how OAM could be simultaneously described by twist-two GTMDs and twist-three GPDs.
Additional relations were discussed, in particular, for the helicity configurations that can be detected analyzing specific spin asymmetries: one corresponding to a longitudinally polarized quark in an unpolarized  proton, associated with spin-orbit correlations, and one for transverse proton polarization, as a generalization of the relation obeyed by the $g_2$ structure function; finally,  we defined a relation connecting the off-forward extension of the Sivers function to an off-forward Qiu-Sterman term.
}

\FullConference{QCD Evolution 2017\\
		22-26 May, 2017\\
		Jefferson Lab Newport News, VA - USA}

\begin{document}
The operator product expansion (OPE) is a powerful tool for analyzing hard scattering processes in QCD.  Within OPE, the twist \cite{Gross:1971wn}, given by the dimension minus the spin of the operators entering the hard processes description, allows one to  order the dominant light cone singularities while separating out their short and long distance components.
At a given value of the hard scale, $Q$, the quark-quark operators of twist two provide the dominant, leading twist terms, while higher twist structure functions are associated with operators of higher dimension and they are suppressed by powers of ${\cal O} (1/Q^{\rm{tw}-2})$. 
Practical evaluations of the hard processes cross section are, however, performed within a  diagrammatic description based on light front quantization.
The way power suppressed terms appear in this approach is through a different construct than OPE:
the quark fields  are decomposed into
"good" and "bad" components, where the good components represent the independent 
terms in the equations of motion \cite{Jaffe:1996zw}. The good components are the leading terms while the bad components are 
suppressed by $1/Q$. The structure functions which contain only good components are of {\em dynamical} twist two. The appearance of bad components introduces terms
of higher dynamical twist.
Dynamical twist is in principle independent from the twist quantum number from OPE. The latter is defined as {\em geometric} since it appears as a consequence of the Lorentz covariance of the theory, without involving the dynamics of the fields. 
Notwithstanding the inherent non-covariance of the light front approach, structure functions of dynamical twist two can be shown to be in one to one correspondence with the geometric twist two terms from OPE.

The underlying difference between the Lorentz invariant structure of OPE and the light front quantization produces a mismatch in the geometric and dynamic twist terms of higher twist.
The Wandzura-Wilczek (WW) relations are a rendition of this mismatch.

The first  WW relation to be derived was the one for the polarized nucleon distribution, $g_2$ \cite{Wandzura:1977qf},
\begin{equation}
\label{WWg2}
g_2(x) = - g_1(x) + \int _x^1 \frac{dy}{y} g_1(y) + \bar{g}_2(x) ,   
\end{equation}
where the dynamical twist three function $g_2$  on the LHS is written in terms of the twist two distribution, $g_1$, and the geometric twist three quark-gluon-quark ($qgq$) correlation, $\bar{g_2}$. 
\footnote{The form of this term will be given below, as well as an additional $qgq$ term originating from the inclusion of the gauge link in the twist two correlation function which is missing from the original formulation.} 
In a nutshell, the physical origin of this relation is from a constraint in the parametrization of the unintegrated correlation functions associated to each specific type of operator, $\hat{O}^\Gamma$, with $\Gamma={\bf 1}, \gamma_5,\gamma^\mu,\gamma^\mu\gamma_5, i \sigma_{\mu \nu}$: the same Lorentz structure combinations that appear in the parametrization with coefficients labeled as $A^i_{E,F,G,H}$ for each operator type ($i$ being an index running over the different types of Lorentz structures), enter simultaneously the corresponding type of dynamical twist two and twist three structure functions obtained integrating over the quark light cone component, $k^-$ \cite{Tangerman:1994bb,Mulders:1995dh}. 

For our specific case, the Parton Distribution Functions (PDFs) $g_1$, $g_T=g_1+g_2$, and the Transverse Momentum Distribution (TMD), $g_{1T}$, are composed of elements of the same set of three amplitudes of the $A_G$ type, and they are therefore connected. The  relation between them  is given by the following equation,
\begin{equation}
\frac{1}{2} \frac{d}{dx} \int d^2 k_T \frac{k_T^2}{M^2} \, g_{1T}(x,k_T^2) = g_T(x)  - g_1(x) - \widehat{g}_T(x)
\end{equation}
This type of relation is called a Lorentz Invariance Relation (LIR). Notice that we included an extra term, $\widehat{g}_T(x)$, originating from  the gauge link ($\widehat{g}_T(x)$ is a so-called LIR violating term \cite{Accardi:2009au,Kanazawa:2015ajw,Goeke:2003az}). 

In a second step, applying the Equations of Motion (EoM) to the correlation function entering the polarized deep inelastic process leads to  a relation involving the same dynamical twist two and three distributions, 
\begin{equation}
 \frac{1}{2}  \int d^2 k_T \frac{k_T^2}{M^2} \, g_{1T}(x,k_T^2) -xg_T (x) +\frac{m}{M} h_1 (x) + x \widetilde{g}_{T} (x) = 0  ,
\end{equation}
where we also included a $qgq$ term, $\widetilde{g}_{T}$, from the covariant derivative  and a term, $m \, h_1$,  from the quark mass appearing  in the EoM.
To obtain a WW relation we now merge the information from the LIR and EoM relation such as to eliminate the TMD moment. We have, 
\begin{equation}
\label{WWg21}
g_T = \int_x^1 \frac{dy}{y} g_1  + \frac{m}{M} \left( \frac{1}{x} h_1 -\int_x^1 \frac{dy}{y^2 } h_1 \right) + \left( \widetilde{g}_{T} -\int_x^1 \frac{dy}{y} \widetilde{g}_{T} \right) +\int_x^1 \frac{dy}{y} \widehat{g}_{T}
\end{equation}
Eq.(\ref{WWg21}) involves twist two and twist three Parton Distribution Functions (PDFs)  plus two distinct $qgq$ terms.
\footnote
{This is a rather new development, see {\it e.g.} Ref.\cite{Accardi:2009au}, at variance with the original formulation (valid for a straight link) in Eq.(\ref{WWg2}). Notice also that the term containing the transversity structure function, $h_1$, originating from the part proportional to quark mass, $m$, in the EoM, disappears in the chiral limit. 
}

The physical content of the geometric twist three functions $\widehat{g}_T$, and $\widetilde{g}_T$, is quite different even if they are both $qgq$ correlations. On one side, for $\widetilde{g}_T$ we have
\cite{Raja:2017xlo},
\begin{equation}
x\widetilde{g}_{T} (x) = \frac{1}{4M} \int d^2 k_T \left( {\cal M}_{+-}^{1,A} +i {\cal M}_{+-}^{2,A} +{\cal M}_{-+}^{1,A} -i {\cal M}_{-+}^{2,A} \right)  \equiv {\cal M}_T^A
\label{gttildedef}
\end{equation}
where,
\begin{eqnarray}
{\cal M}^{i, S}_{\Lambda^{\prime} \Lambda } \! &=& \! -\frac{i}{4} \int \frac{d z^- d^2 z_T}{(2 \pi)^3} e^{ixP^+ z^- - i k_T\cdot z_T}  \nonumber \\
& \times & \langle p^{\prime } ,\Lambda^{\prime } \mid \overline{\psi} \left(-\frac{z}{2} \right) \left[
\left. (\overrightarrow{\slashed{\partial } } -ig\slashed{A} )
{\cal U} \Gamma \right|_{-z/2} 
 +   \left. \Gamma {\cal U}
(\overleftarrow{\slashed{\partial } } +ig\slashed{A} ) \right|_{z/2}
\right] \psi \left(\frac{z}{2} \right) \mid p, \Lambda \rangle_{z^+=0} \nonumber \\ 
\label{qgqterms1} \\
{\cal M}^{i, A}_{\Lambda^{\prime } \Lambda } \! &=& \! -\frac{i}{4}
\int \frac{d z^- d^2 z_T}{(2 \pi)^3} e^{ixP^+ z^- - i k_T\cdot z_T} \nonumber \\
& \times & \langle p^{\prime } ,\Lambda^{\prime } \mid \overline{\psi} \left(-\frac{z}{2} \right) \left[
\left. (\overrightarrow{\slashed{\partial } } -ig\slashed{A} )
{\cal U} \Gamma \right|_{-z/2} - \left. \Gamma {\cal U}
(\overleftarrow{\slashed{\partial } } +ig\slashed{A} ) \right|_{z/2}
\right] \psi \left( \frac{z}{2} \right)  \mid p, \Lambda \rangle_{z^+=0} \nonumber \\ 
\label{qgqterms}
\end{eqnarray}
where $\Lambda (\Lambda')$ are the helicities for the proton in its initial and final states, respectively. 

On the other hand, $\widehat{g}_T$, is defined as \cite{Raja:2017xlo},
\begin{equation}
\widehat{g}_{T} (x) = \frac{d}{dx} \left( {\cal M}_T^A - \left. {\cal M}_T^A \right|_{v=0} \right) 
\label{gthatdef}
\end{equation}
To ensure gauge invariance, the quark bilocal operator in the correlation function requires
a gauge link, ${\cal U}$ along a path connecting the quark operator positions
$-z/2$ and $z/2$. Two important choices of path are a direct
straight line and a staple-shaped connection characterized by an additional
four-vector $v$. $v$ describes the legs of the staple-shaped
path. In the special case $v=0$,  the
staple degenerates to a straight link between the two positions.
The different path choices will give rise to different geometric twist three contributions to the correlators.

The detailed twist three contributions to polarized structure functions were studied in Ref.\cite{Accardi:2009au} where  the numerical difference between $\widetilde{g}_T$ and $\widehat{g}_T$ was found to be undetermined in phenomenological extractions from data.

\vspace{0.5cm}
In Refs.\cite{Raja:2017xlo,Rajan:2016tlg} we showed how these effects impact the off-forward case in a more fundamental way since they turn out to be associated to the physical interpretation of Orbital Angular Momentum (OAM). Through deeply virtual exclusive experiments one can study the spin correlations for an unpolarized quark in a longitudinally polarized proton ($UL$), or vice-versa for the polarized quark inside an unpolarized proton ($LU$). As shown in \cite{Liuti:2017uxp}, these correlations measuring twist three GPDs are associated to specific azimuthal angular dependent terms in the cross section.

The LIRs involving twist three GPDs and twist two GTMDs relevant to our case are,
for the vector and axial vector cases respectively,   
\begin{eqnarray}
\label{eq:LIRviolf14}
\frac{d}{dx}   \int d^2 k_T \frac{k_T^2}{M^2} F_{14}  & = & \widetilde{E}_{2T} +H +E + {\cal A}_{F_{14} } \\
\label{eq:LIRviolg11}
\frac{d}{dx}    \int d^2 k_T \frac{k_T^2}{M^2} G_{11} & = & -\left(2\widetilde{H}_{2T}' + E_{2T}'
 \right) - \widetilde{H}+ {\cal A}_{G_{11} }  \, ,
 \end{eqnarray}
where the $k_T$-integrals over the twist two vector and axial-vector GTMDs, $F_{14}$ and  $G_{11}$,  can be associated, respectively, with the $x$ distributions of the longitudinal OAM component, $L_z$,  and the quark spin-orbit, $L_z S_z$  \cite{Lorce:2011kd}; the combinations,  $ \widetilde{E}_{2T} +H +E $, and $-(2\widetilde{H}_{2T}' + E_{2T}') - \widetilde{H}$ are expressed in an analogous format as shown for $g_T=g_1+g_2$, $H +E$ and $\widetilde{E}_{2T}$ being vector twist two and twist three GPDs, in this case, and similarly, $\widetilde{H}$ and $2\widetilde{H}_{2T}' + E_{2T}'$ being axial-vector twist two and twist three GPDs; ${\cal A}_{F_{14} } $ and ${\cal A}_{G_{11} }$ are LIR violating terms analogous to $\widehat{g}_T$, (Eq.(\ref{gthatdef}), that can be expressed in terms of genuine twist three contributions.
These relations are valid point by point in the kinematical variables $x$ and $t=\Delta^2$ ($\Delta=p'-p$, being the momentum transfer between the initial and final protons in the deeply virtual exclusive scattering process). 

The EoM relations for the off-forward case can be written as \cite{Raja:2017xlo,Rajan:2016tlg}, 
\vspace{0.5cm}
\begin{eqnarray}
\label{eq:EoMF14}
x \widetilde{E}_{2T}(x)  &=& - \widetilde{H}(x) +  \int d^2 k_T \frac{k_T^2}{M^2} F_{14}  - {\cal M}_{F_{14} }  \\
\label{eq:EoMG11}
x\left[2\widetilde{H}_{2T}'(x) + E_{2T}'(x) \right] & = &- H(x)  + \frac{m}{M}( 2 \widetilde{H}_T(x)+ E_T(x))  - \int d^2 k_T \frac{k_T^2}{M^2} G_{11} - {\cal M}_{G_{11} }
\end{eqnarray}
where the $qgq$ term now read, 
\begin{eqnarray}
&& \mathcal{M}_{F_{14}} (x) = \int d^2 k_T \, \frac{\Delta^{i} }{\Delta_T^2} \left( \mathcal{M}^{i,S}_{++} - \mathcal{M}^{i,S}_{--} \right) \\
&& \mathcal{M}_{G_{11}} (x) = i\epsilon^{ij} \int d^2 k_T  \, \frac{\Delta^{j} }{\Delta_T^2} \left( {\mathcal{M}}^{i,A}_{++} + {\mathcal{M}}^{i,A}_{--} \right) \quad ,
\end{eqnarray}
with the expressions for ${\mathcal{M}}^{i,S(A)}_{\Lambda \Lambda'}$  given in Eqs.(\ref{qgqterms1},\ref{qgqterms}).

Eliminating the GTMD term from Eqs.(\ref{eq:LIRviolf14},\ref{eq:LIRviolg11}) and (\ref{eq:EoMF14},\ref{eq:EoMG11}), we obtain the following off-forward WW relations, 
\begin{eqnarray}
\label{F14_WW1}
&& \widetilde{E}_{2T} =
- \int_x^1 \frac{dy}{y}(H + E) - \left[ \frac{\widetilde{H}}{x} -\int_x^1 \frac{dy}{y^2} \widetilde{H}\right]  - \left[ \frac{1}{x}\mathcal{M}_{F_{14}} - \int_x^1 \frac{dy}{y^2} \mathcal{M}_{F_{14}}  \right]
- \int_x^1 \frac{dy}{y} {\cal A}_{F_{14} }
\\
\label{G11_WW1}
&& 2\widetilde{H}_{2T}' + E_{2T}' = 
-\int_x^1 \frac{dy}{y} \widetilde{H} - \left[  \frac{H}{x} - \int_x^1 \frac{dy}{y^2} H \right] + \frac{m}{M}\left[ \frac{1}{x}(2 \widetilde{H}_T + E_T ) - \int_x^1 \frac{dy}{y^2}(2 \widetilde{H}_T + E_T )\right] 
\nonumber \\
&& \quad \quad \quad - \left[\frac{1}{x} \mathcal{M}_{G_{11}} -\int_x^1 \frac{dy}{y^2} \mathcal{M}_{G_{11}} \right] +\int_x^1 \frac{dy}{y} {\cal A}_{G_{11} }
\end{eqnarray}
Eqs.(\ref{eq:LIRviolf14}-\ref{G11_WW1}) are valid for either a staple or a straight gauge link structure (with staple vector $v$ on the light cone in the former case), keeping in mind that ${\cal A}_{F_{14} } \equiv 0$ and ${\cal A}_{G_{11} } \equiv 0$ in the straight-link case. 


The gauge link structure takes on a specific meaning for OAM. It is well known that an ambiguity is introduced in performing the decomposition of the total angular momentum, $J$, as defined from the QCD energy momentum tensor, into its quark and gluon $L$ and $S$ components. In the quark sector, in particular, the two decompositions known as Jaffe-Manohar (JM) and Ji, differ from one other in the definition of $L$, in that they involve a partial and a covariant derivative, respectively, thus implying different contributions from the gluonic component. In Refs. \cite{Hatta:2011ku,Hatta:2012cs,Burkardt:2012sd} this contribution was worked out explicitly; in particular in Ref.\cite{Burkardt:2012sd}  it was shown that JM OAM is calculated using a staple link, while Ji OAM uses a straight link, or a link with length $v=0$. In Ref.\cite{Raja:2017xlo} we found a concrete expression for this difference, in terms of the invariant amplitudes, $A^F_i$, listed in Ref.\cite{Meissner:2009ww}, namely
 \begin{eqnarray}
{\cal A}_{F_{14} } (x) & \equiv &
v^{-} \frac{(2P^{+} )^2 }{M^2 } \int d^2 k_T \int dk^{-} \left[
\frac{k_{T} \cdot \Delta_{T} }{\Delta_{T}^{2} } (A^F_{11} +xA^F_{12} )
+A^F_{14} \right. \nonumber \\
& + & \left. \frac{k_T^2 \Delta_{T}^{2} - (k_T \cdot \Delta_{T} )^2 }{\Delta_{T}^{2} } \left(
\frac{\partial A^F_8 }{\partial (k\cdot v)} +
x\frac{\partial A^F_9 }{\partial (k\cdot v)} \right) \right]  = \nonumber \\
&=& \frac{d}{dx}  \int d^2 k_T \frac{k_T^2}{M^2} F_{14} -
\left. \frac{d}{dx}  \int d^2 k_T \frac{k_T^2}{M^2} F_{14}  \right|_{v=0} = \frac{d}{dx} \left(
{\cal M}_{F_{14} } - \left. {\cal M}_{F_{14} } \right|_{v=0} \right) .
\label{af14diff}
\end{eqnarray}
We can reexpress the relation above in order to emphasize its description of OAM as,
\begin{eqnarray}
  L^{JM}(x) - L^{Ji}(x) =  
{\cal M}_{F_{14} } - \left. {\cal M}_{F_{14} } \right|_{v=0}  = - \int_x^1  dy \, {\cal A}_{F_{14} }(y)  .
\label{JMJi}
\end{eqnarray}
(we likewise obtains the analogous relation for ${\cal A}_{G_{11} } $). 

We conclude that an experimental determination of the difference between JM and Ji OAM, described by ${\cal A}_{F_{14} } $, is possible only by measuring separately the GPDs $\widetilde{E}_{2T}$, $H$ and $E$, and the GTMD $F_{14}$, using Eq.(\ref{eq:LIRviolf14}). GTMD measurements are much harder because of the number of variables involved: the proposed reactions, to date, can be considered only at a speculative level \cite{Hagiwara:2017fye,Bhattacharya:2017bvs}. GTMDs can be, however, evaluated in lattice QCD calculations \cite{Engelhardt:2017miy}. Notwithstanding, an important spin-off from our study is that we solve the question of an ambiguity in the composition of the intrinsic $qgq$ twist three contribution to the twist three structure functions raised in Ref.\cite{Accardi:2009au}. Assuming that all GPDs, being collinear objects, are not affected by the gauge link path, from Eq.(\ref{JMJi}) we see that the $qgq$ term for a staple link which is composed by two types of contributions (from ${\cal M}$ and ${\cal A}$), must be equal to the  $qgq$ term for a straight link, which is composed by only one term (${\cal M}$). Experimental measurements of twist three GPDs and PDFs do not allow us to distinguish among the two.


\begin{thebibliography}{99}
\bibitem{Gross:1971wn} 
  D.~J.~Gross and S.~B.~Treiman,
  Phys.\ Rev.\ D {\bf 4}, 1059 (1971).
\bibitem{Jaffe:1996zw} R.~L.~Jaffe,
  ``Spin, twist and hadron structure in deep inelastic processes,''
 hep-ph/9602236.
  \bibitem{Wandzura:1977qf}S.~Wandzura and F.~Wilczek,
  Phys.\ Lett.\  {\bf 72B}, 195 (1977).
\bibitem{Tangerman:1994bb} R.~D.~Tangerman and P.~J.~Mulders,
  hep-ph/9408305.
  \bibitem{Mulders:1995dh}  P.~J.~Mulders and R.~D.~Tangerman,
   Nucl.\ Phys.\ B {\bf 461}, 197 (1996)
  Erratum: [Nucl.\ Phys.\ B {\bf 484}, 538 (1997)]
\bibitem{Accardi:2009au} A.~Accardi, A.~Bacchetta, W.~Melnitchouk and M.~Schlegel,
  JHEP {\bf 0911}, 093 (2009)
\bibitem{Kanazawa:2015ajw} 
  K.~Kanazawa, Y.~Koike, A.~Metz, D.~Pitonyak and M.~Schlegel,
  Phys.\ Rev.\ D {\bf 93}, no. 5, 054024 (2016)
\bibitem{Goeke:2003az} 
  K.~Goeke, A.~Metz, P.~V.~Pobylitsa and M.~V.~Polyakov,
  Phys.\ Lett.\ B {\bf 567}, 27 (2003)
\bibitem{Raja:2017xlo} 
  A.~Rajan, M.~Engelhardt and S.~Liuti,
  arXiv:1709.05770 [hep-ph].
\bibitem{Rajan:2016tlg} 
  A.~Rajan, A.~Courtoy, M.~Engelhardt and S.~Liuti,
  Phys.\ Rev.\ D {\bf 94}, no. 3, 034041 (2016)
\bibitem{Liuti:2017uxp} 
  S.~Liuti, "{Workshop on High-Intensity Photon Sources (HIPS2017)
                        Mini-Proceedings}", 2017.
 \bibitem{Lorce:2011kd}
C.~Lorce and B.~Pasquini,
  Phys.\ Rev.\ D {\bf 84}, 014015 (2011)
\bibitem{Hatta:2011ku} 
  Y.~Hatta,
  Phys.\ Lett.\ B {\bf 708}, 186 (2012)
\bibitem{Hatta:2012cs} 
  Y.~Hatta and S.~Yoshida,
  JHEP {\bf 1210}, 080 (2012)
\bibitem{Burkardt:2012sd} 
M.~Burkardt,
  Phys.\ Rev.\ D {\bf 88}, no. 1, 014014 (2013)
\bibitem{Meissner:2009ww} 
  S.~Meissner, A.~Metz and M.~Schlegel,
  JHEP {\bf 0908}, 056 (2009)
\bibitem{Hagiwara:2017fye} 
  Y.~Hagiwara, Y.~Hatta, R.~Pasechnik, M.~Tasevsky and O.~Teryaev,
  Phys.\ Rev.\ D {\bf 96}, no. 3, 034009 (2017)
\bibitem{Bhattacharya:2017bvs} 
  S.~Bhattacharya, A.~Metz and J.~Zhou,
  Phys.\ Lett.\ B {\bf 771}, 396 (2017)
\bibitem{Engelhardt:2017miy} 
  M.~Engelhardt,
  Phys.\ Rev.\ D {\bf 95}, no. 9, 094505 (2017)
     
\end{thebibliography}
\end{document}